\documentclass[11pt]{article}

\usepackage{latexsym,amsmath,epsfig}

\DeclareFontFamily{OT1}{rsfs10}{}
\DeclareFontShape{OT1}{rsfs10}{m}{n}{ <-> rsfs10 }{}
\DeclareMathAlphabet{\mathscript}{OT1}{rsfs10}{m}{n}

\numberwithin{equation}{section}

\newcommand{\be}{\begin{equation}}
\newcommand{\ee}{\end{equation}}
\newcommand{\nn}{\nonumber}
\newcommand{\bea}{\begin{eqnarray}}
\newcommand{\eea}{\end{eqnarray}}

\newcommand{\tr}{\textrm{tr}}
\newcommand{\ns}{\normalsize}
\newcommand{\pt}{\partial}

\newcommand{\mbf}[1]{\mathbf{#1}}

\def\a{\alpha}
\def\b{\beta}
\def\g{\gamma}
\def\c{\chi}
\def\d{\delta}
\def\e{\epsilon}

\def\f{\phi}

\def\z{\psi}

\def\k{\kappa}
\def\l{\lambda}
\def\m{\mu}
\def\n{\nu}
\def\o{\omega}
\def\p{\pi}

\def\r{\rho}
\def\s{\sigma}

\def\t{\tau}

\def\x{\xi}
\def\z{\zeta}

\def\G{\Gamma}
\def\J{\Psi}
\def\L{\Lambda}
\def\O{\Omega}

\def\cA{{\cal A}}
\def\cB{{\cal B}}
\def\cF{{\cal F}}
\def\cH{{\cal H}}

\def\cN{{\cal N}}
\def\cK{{\cal K}}

\def\Ib{\bar{I}}
\def\Jb{\bar{J}}
\def\Kb{\bar{K}}
\def\Lb{\bar{L}}

\def\bbar{\bar{b}}

\def\RR{\mbox{Re}}

\def\Tb{\bar{T}}
\def\Sb{\bar{S}}
\def\Fb{\bar{F}}
\def\nb{\bar{n}}
\def\mb{\bar{m}}

\begin{document}


\begin{titlepage}

\vspace{-3cm}

\title{
   \hfill{\ns OUTP-98-93P, UPR-828T, PUPT-1829\\}
   \hfill{\ns hep-th/9901017} \\[3em]
   {\huge Five--Branes and Supersymmetry Breaking in M--Theory}\\[1em]}
\author{
{\ns\large Andr\'e Lukas$^1$, Burt A.~Ovrut$^2$
            and Daniel Waldram$^3$} \\[0.8em]
   {\it\ns $^1$Department of Physics, Theoretical Physics, 
     University of Oxford} \\[-0.5em]
      {\it\ns 1 Keble Road, Oxford OX1 3NP, United Kingdom} \\[0.5em]   
   {\it\ns $^2$Department of Physics, University of Pennsylvania} \\[-0.5em]
      {\it\ns Philadelphia, PA 19104--6396, USA}\\[0.5em]
   {\it\ns $^3$Department of Physics, Joseph Henry Laboratories,} \\[-0.5em]
      {\it\ns Princeton University, Princeton, NJ 08544, USA}}
\date{}

\maketitle

\begin{abstract}
Supersymmetry breaking via gaugino condensation is studied in vacua of
heterotic M--theory with five--branes. We show that supersymmetry is
still broken by a global mechanism and that the non--perturbative
superpotential takes the standard form. When expressed in terms of low
energy fields, a modification arises due to a threshold correction in
the gauge kinetic function that depends on five--brane moduli. We also
determine the form of the low energy matter field K\"ahler potential.
These results are used to discuss the soft supersymmetry breaking
parameters, in particular the question of universality.
\end{abstract}

\thispagestyle{empty}

\end{titlepage}


\section{Introduction}

Starting with the seminal work by Ho\v rava and
Witten~\cite{hw1,hw2,hor,w}, considerable
activity~\cite{bd}--\cite{bkl1} has been
devoted to exploring various aspects of M--theory on the
orbifold $S^1/Z_2$ which represents the strong coupling limit of
$E_8\times E_8$ heterotic string theory. Most of the work dealing with
compactifications of this theory based on Calabi--Yau three--folds has
used the standard--embedding of the spin connection into the gauge
group as a starting point. Although this leads to particularly simple
solutions of the weakly coupled heterotic string, this is not the case
for its strongly coupled counterpart. In fact, there does not seem
to be any way, for Calabi--Yau type compactifications, to set the
antisymmetric tensor field of 11--dimensional supergravity to zero.
Hence, the standard--embedding does not appear to be particularly
special in the strong coupling limit and one is led to consider the
general case of non--standard embeddings. Aspects of such non--standard
embeddings, in vacua of heterotic M--theory not involving five--branes
have been discussed in~\cite{benakli1,stieb,lpt,nse}.

It turns out that such non--standard embedding vacua are, in fact, a small
subclass of a much larger class of vacua that, in addition, allows for the
presence of five--branes. Including five--branes is
quite natural from the M--theory viewpoint once general non--standard
embeddings are allowed and is essential for a ``complete''
discussion of such vacua. These non--perturbative vacua, based on Calabi--Yau
three--folds, general $E_8$ gauge bundles and a number of five--branes,
have been constructed in two recent papers~\cite{nse,np}. The
five--branes are oriented such that they stretch across the
$3+1$--dimensional uncompactified space and wrap on two--cycles within
the Calabi--Yau manifold. 

These non--perturbative vacua have a number of new and perhaps surprising
properties. Most notably, it has been explicitly shown~\cite{np} that
three--generation models with attractive low--energy GUT groups exist for
most Calabi--Yau spaces in a certain specific class. This is
facilitated by the presence of the five--branes which introduce
considerably more freedom in the anomaly cancellation condition that
consistent vacua need to satisfy. These results suggest that
three--generation models exist for most Calabi--Yau spaces once
five--branes are allowed in the vacuum. Another interesting feature is
the appearance of new sectors in the low energy theory that originate
from the degrees of freedom on the five--brane worldvolumes. In
particular, these new sectors carry Abelian or non--Abelian gauge
theories that enlarge the total low energy gauge group. At low energy,
all these sectors are coupled to one another and to the two
conventional sectors that arise from $E_8\times E_8$ only
gravitationally. Another interesting feature is that the two $E_8$
sectors are modified due to the presence of the five--branes. For
example, the gauge kinetic functions depend on the moduli which
specify the positions of the five--branes in the orbifold
direction. The appearance of these moduli is a 
non--perturbative effect that cannot be seen in the weakly coupled
limit of the heterotic string.  These properties make vacua of
heterotic M--theory with five--branes a rich and interesting arena
for particle physics.

In this paper, we would like to investigate some of the physical
properties of these new non--perturbative vacua. Specifically, we
will analyze in detail how
supersymmetry breaking via gaugino condensation~\cite{din,drsw} in the
hidden sector works. For standard embedding vacua, gaugino
condensation in heterotic M--theory was first analyzed by
Ho\v rava~\cite{hor} and subsequently by~\cite{lt,gau}. Supersymmetry
breaking in five dimensions was also studied in~\cite{mp,elpp,elpp1}.
Furthermore, we will calculate the matter field
part of the $D=4$ low energy effective action in its generic form.
These results will then be used to discuss soft supersymmetry breaking
terms and, in particular, the question of universality. 
Recently, supersymmetry breaking and universality of the soft breaking
parameters in supergravity theories that originate from vacua with branes has
also been discussed in ref.~\cite{rs}. 

It is important to be clear about the interpretation of the vacua from
the low energy viewpoint. In this paper, we will use a
``conservative'' interpretation in that we assume the observable
sector to arise from one of the $E_8$ sectors. Consequently, the
five--brane sectors, as well as the other $E_8$ sector, are hidden. In
addition, we assume that supersymmetry breaking via gaugino
condensation occurs in the hidden $E_8$ sector. These assumptions
could be modified in two ways. First, it is conceivable that, under
certain conditions, one of the five--brane sectors could also be used
as the observable sector. Although models along those lines have yet
to be constructed in heterotic M--theory, there are recent type I
examples~\cite{ibanez,lykken,biq} where semi--realistic low--energy
theories arise
from three--brane worldvolume theories. In any case, the relation of
our constructions to brane--box models~\cite{hz,hu} suggests that chiral
theories from the five--brane sectors might exist. As a second
modification, gaugino condensation could also occur in those hidden
five--brane sectors that carry non--Abelian gauge groups. Although
these modifications are worth investigating, in this paper we stick
to the simple interpretation explained above. In the end, we briefly
comment on gaugino condensation in the five--brane sectors. 

An important question we would like to address is how the presence of
the five--branes affects supersymmetry breaking via gaugino
condensation. For the standard--embedding case, Ho\v rava has shown
that supersymmetry is broken by a mechanism that is global in the orbifold
direction. Despite this new feature, the non--perturbative low energy
potential induced by this condensate turned out to be the same as in
the weakly coupled case~\cite{bd,gau}. The five--branes present in our
non--perturbative vacua are stacked up along the orbifold direction;
that is, precisely the direction in which the global breaking
mechanism occurs. It is, therefore, quite conceivable that the five--branes can
significantly alter the picture of supersymmetry breaking via gaugino
condensation. 

To analyze this in detail, we will first review the form of the
vacua with five--branes, both from a topological and an analytic
viewpoint. In particular, we present the explicit form of those vacua
as an expansion, to linear order, in the strong coupling expansion
parameter $\e =(\k /4\p )^{2/3}2\p^2\r /v^{2/3}$. Here $\k$ is the
eleven--dimensional Newton constant, $v$ is the Calabi--Yau volume and
$\r$ is the orbifold radius. Subsequently, we incorporate gaugino
condensation and analyze how supersymmetry is broken on the level of
the vacuum solution. In section 3, we compute the $D=4$ superpotential
from gaugino condensation in terms of the condensate. While this was
done in the 11--dimensional theory, in section 4, we present an
alternative derivation in the five--dimensional effective theory of
heterotic M--theory. To express the superpotential in terms of low
energy fields, we need to compute the matter field part of the
effective four--dimensional theory which is done in section 5.
Finally, in section 6, we use these results to discuss soft
supersymmetry breaking terms.

Our results can be summarized as follows. On the level of the pure
vacuum solution, supersymmetry breaking via gaugino condensation in
the presence of five--branes still occurs via Ho\v rava's global
mechanism. Furthermore, the condensate superpotential as expressed in
terms of the condensate is unchanged with respect to the ordinary
case. Roughly speaking, these results can be attributed to the fact that the
gaugino condensate is associated with a different sector of the
Calabi--Yau space than the curves on which the five--branes are
wrapped. Once the non--perturbative condensate potential is expressed
in terms of the moduli fields, a new feature related to the presence of
five--branes occurs. It turns out that the gauge kinetic functions in
the $E_8$ sectors have threshold corrections of order $\e $ which
not only depend on the $T$ moduli, but also on the moduli that specify
the positions of the five--branes in the orbifold direction~\cite{nse}.
Hence, the potential not only depends on the dilaton $S$ and the
$T$--moduli $T^i$, but also on the five--brane position moduli $Z^n$. 
A similar correction of order $\e $ to the matter field K\"ahler
metric depending on $S$ and $Z^n$ is found as well. 
Based on these results, we propose a parameterization of supersymmetry
breaking in terms of the auxiliary fields associated to $S$,
$T^i$ and the five--brane moduli $Z^n$. Using this parameterization, we
analyze the question of universality of soft parameters.
Universality can be achieved at zeroth order in $\e $
by assuming supersymmetry breaking in the $S$ or $Z^n$ direction.
As usual, breaking in the $T^i$ directions is not
generation--independent, even at zeroth order. From an
11--dimensional point of view this happens because both the
$T$--moduli and the matter fields arise as zero modes of the
Calabi--Yau space. When the order $\e $ corrections to the matter field
K\"ahler metric are taken into account no specific supersymmetry
breaking pattern leads to universality. Generally, not even a pure $S$--type
breaking scenario~\cite{kl} will work. The 11--dimensional reason for
this is an orbifold dependence of the background on which
the reduction to four dimensions is performed. This dependence arises
from the sources provided by the fields on the orbifold planes and the
five--branes and, as such, is a general phenomenon for vacua containing
branes. We observe that this difficulty can be overcome by choosing a
specific type of
Calabi--Yau space with $h^{1,1}=1$. In this case, universality to
order $\e $ is guaranteed independently of the breaking pattern.
Soft breaking parameters are then computed for this particular case. 
These parameters can be of the order $m_{3/2}$ or $\e m_{3/2}$ with a
characteristic signature that is directly correlated to the pattern of
supersymmetry breaking.

\section{The eleven--dimensional background}

In this section, we would like to review the vacua of heterotic
M--theory with non--standard embedding and five--branes constructed
in ref.~\cite{nse,np}. In addition, we would like to incorporate
gaugino condensates on the orbifold planes and show, on the level of
the vacuum solution, how supersymmetry is broken by those condensates.
Most importantly, we would like to understand the role of the
five--branes in this context. None of the ingredients needed for this
is essentially new. Nevertheless we find it useful to
present a coherent argument and to set the stage for what is going to
follow.

Let us first explain the basic structure of the non--perturbative
vacua and their topological properties following ref.~\cite{np}.
To lowest order, we consider a space--time
\begin{equation}
   M_{11}=M_{4} \times S^{1}/Z_{2} \times X
\label{spacetime}
\end{equation}
where $X$ is a Calabi--Yau three--fold and $M_4$ is four--dimensional
Minkowski space. Furthermore, one needs to specify two semi--stable
gauge bundles $V_i$, $i=1,2$ on $X$ with gauge groups
$G_i\subset E_8$. As discussed, we also allow for the presence of
$N$ five--branes in the vacuum. These five--branes should
stretch across $M_4$ (to preserve $3+1$--dimensional Poincar\'e
invariance) and wrap on holomorphic curves $C^{(n)}\subset X$ where
$n=1,\dots ,N$ (to preserve $\cN =1$ supersymmetry in four
dimensions). Note that, given this
orientation, the five--branes are located at specific points in the
orbifold direction which we denote by $x^{11}=x_1,\dots ,x_N$, where
$0\leq x_1 \leq\cdots\leq x_N\leq\p\r$. As we will discuss later on,
inclusion of five--branes is important as it allows for the
construction of physically interesting vacua that would otherwise
be forbidden. Heuristically, this can be understood from the cohomology
constraint
\begin{equation}
   c_{2}(V_1) + c_{2}(V_2) + [W] = c_{2}(TX)
\label{cohcon}
\end{equation}
that has to be imposed on such vacua in order to be anomaly free. Here
$[W]$ is the cohomology class associated with the five--branes
and $c_{2}(V_{i})$ and $c_{2}(TX)$ are the second Chern classes of
the gauge bundles $V_{i}$ and the tangent bundle $TX$ respectively.
In fact, for a given Calabi--Yau space and given vector bundles one
can always choose a class $[W]$ such that eq.~\eqref{cohcon} is
satisfied. As was shown in ref.~\cite{np}, there is an additional
physical constraint on $[W]$, namely that it constitutes an effective
class. Essentially, this means that $[W]$ corresponds to holomorphic
curves in $X$; that is, the associated five--brane must be of the form
$W=\sum_{n=1}^NC^{(n)}$ where the $C^{(n)}$ are linear combinations of
holomorphic curves with non--negative integer coefficients. This has
to be contrasted to the situation without five--branes where simply
$[W]=0$.

In ref.~\cite{np}, these statements have been made much more precise by
constructing $SU(n)$ gauge bundles over elliptically fibered
Calabi--Yau spaces. This allows one to explicitly compute the
Chern--classes $c_2(TX)$ and $c_2(V_i)$ and hence, from
eq.~\eqref{cohcon}, the class $[W]$ of the five--brane curve. It was
shown that the effectiveness constraint can be satisfied and
further physical requirements can be imposed. In particular, a
three--generation constraint can be accommodated without problems
along with, for example, an $SU(5)$ gauge bundle that leads to a low energy
$SU(5)$ grand unification group. It was also demonstrated, for a
specific example, how to determine the moduli space of curves $W$
associated with the class $[W]$. This moduli space turns out to be
a complicated multi--branched object that carries information about
the additional five--branes sectors. In particular, sticking to a
specific branch, it allows one to read off the number of five--branes,
$N$, and the classes $[C^{(n)}]$ of their associated curves. With
this information one can define the instanton numbers $\b_i^{(0)}$,
$\b_i^{(N+1)}$ and the five--brane charges $\b_i^{(n)}$, $n=1,\dots ,N$ by
\bea
 c_2(V_1)-\frac{1}{2}c_2(TX) &=& \b_i^{(0)}C_2^i\nn \\
 c_2(V_2)-\frac{1}{2}c_2(TX) &=& \b_i^{(N+1)}C_2^i\label{instanton}\\
 \left[ C^{(n)}\right] &=& \b_i^{(n)}C_2^i\nn\; .
\eea
Here $\{ C_2^i\}$ is a basis of the second homology group $H_2(X,\mbf{Z})$
(which we identify with $H^4(X,\mbf{Z})$ via Poincar\'e duality). The
cohomology constraint~\eqref{cohcon} then
takes the form
\begin{equation}
 \sum_{n=0}^{N+1}\b_i^{(n)} = 0
\end{equation}
for all $i=1,\dots ,h^{1,1}$. The numbers $\b_i^{(n)}$ constitute the
essential topological information that appears in the explicit form of
the vacua.

Let us now review this explicit form following
ref.~\cite{nse}. Generally, an 11--dimensional M--theory vacuum is
specified by the metric $g_{IJ}$ and the three--form $C_{IJK}$ with field
strength $G_{IJKL}=24\,\partial_{[I}C_{JKL]}$. For our discussion of
supersymmetry breaking, we will also need the Killing spinor $\eta$
corresponding to four preserved supercharges. We will use
11--dimensional indices $I,J,K,\dots = 0,\dots ,9,11$,
four--dimensional indices $\m ,\n ,\r ,\dots = 0,\dots 3$ and
Calabi--Yau indices $A,B,C,\dots =4,\dots ,9$. Holomorphic and
anti--holomorphic indices on the Calabi--Yau space are denoted by
$a,b,c,\dots$ and $\bar{a},\bbar ,\bar{c},\dots$.
The form of the vacuum has been computed using an expansion, to linear
order, in the parameter
\begin{equation}
 \e  = \left(\frac{\k}{4\p}\right)^{2/3}
        \frac{2\p^2\r}{v^{2/3}}\label{es}\; .
\end{equation}
Here $\k$ is the 11--dimensional Newton constant, $v$ is the
Calabi--Yau volume and $\r$ is the orbifold radius.
To this order the solution takes the form~\cite{w}
\bea
 ds^2 &=& (1+b)\eta_{\m\n}dx^\m dx^\n +(g^{({\rm CY})}_{AB}+
          h_{AB})dx^Adx^B+(1+\g )(dx^{11})^2 \nn \\
 G_{ABCD} &=& G^{(1)}_{ABCD}\nn \\
 G_{ABC11} &=& G^{(1)}_{ABC11} \label{solform}\\
 \eta &=& (1+\psi )\eta^{({\rm CY})}\nn
\eea
and all other components of $G$ vanishing.
Here $g^{({\rm CY})}_{AB}$ and $\eta^{({\rm CY})}$ are the Ricci--flat
metric and the covariantly constant spinor on the Calabi--Yau space.
The quantities $b$, $h_{AB}$, $\g$, $G^{(1)}$ and $\psi$ represent the
corrections linear in $\e $. Following ref.~\cite{low} they can be
expressed in terms of a single $(1,1)$--form $\cB_{a\bbar}$ on $X$ as follows
\begin{equation}
\begin{aligned}
  h_{a\bbar} &= \sqrt{2}i \left( \cB_{a\bbar} 
     - \frac{1}{3}\o_{a\bbar}\cB \right) \\
  b &= \frac{\sqrt{2}}{6} \cB  \\
  \g &= -\frac{\sqrt{2}}{3} \cB \\
  \psi &= -\frac{\sqrt{2}}{24} \cB  \\
  G_{ABCD}^{(1)} &= \frac{1}{2}\e_{ABCDEF}\partial_{11}\cB^{EF}  \\
  G_{ABC11}^{(1)} &= \frac{1}{2}\e_{ABCDEF}\partial^D\cB^{EF}\; .
\end{aligned}
\label{sol} 
\end{equation}
This $(1,1)$--form can be expanded in terms of harmonics of the
Calabi--Yau Laplacian. For the purpose of computing low energy
effective actions, it is sufficient to keep the massless terms in this
expansion; that is, the terms proportional to the harmonic
$(1,1)$--forms of the Calabi--Yau space. Let us choose a basis
$\{\o_{ia\bbar}\}$ of those forms dual to the cycles
$C_2^i$ introduced above. We then write
\begin{equation}
 \cB_{a\bbar} = \sum_ib_i\o_{a\bbar}^i+(\mbox{massive terms})\label{bser}
\end{equation}
where the indices are lowered and raised with the metric
\begin{equation}
 G_{ij} = \frac{1}{2v}\int_X\o_i\wedge (*\o_j)\label{metric11}
\end{equation}
on the K\"ahler moduli space. The expansion coefficients $b_i$ are
given by~\cite{nse}
\begin{equation}
 b_i = \frac{\e }{\sqrt{2}}\left[\sum_{m=0}^n
       \b_i^{(m)}(|z|-z^m)-\frac{1}{2}\sum_{m=0}^{N+1}(1-z^m)^2
       \b_i^{(m)}\right]
\label{massless}
\end{equation}
in the interval
\begin{equation}
 z^n\leq |z|\leq z^{n+1}\; ,
\end{equation}
for fixed $n$, where $n=0,\dots ,N$. Here $z=x^{11}/\p\r\in [-1,1]$ is
the normalized orbifold coordinate and $z^n=x_n/\p\r\in [0,1]$ where
$n=1,\dots ,N$ are the five--brane positions in this coordinate
frame. We have also defined $z_0=0$ and $z_{N+1}=1$. To summarize,
eqs.~\eqref{solform}--\eqref{massless} determine the background to linear
order in $\e $. As promised, the only input needed are the topological
numbers $\b_i^{(n)}$ defined in eq.~\eqref{instanton}.
As we have discussed above, they can computed by methods of algebraic
geometry, at least for certain classes of Calabi--Yau spaces.

\vspace{0.4cm}

Let us now analyze the properties of these vacua in the presence of
gaugino condensates. We recall that, following Ho\v rava~\cite{hor},
the proper way to avoid singularities in the presence of condensates
is to redefine the antisymmetric tensor field according to
\bea
 \tilde{G}_{\bar{I}\bar{J}\bar{K}\bar{L}}
     &=& G_{\bar{I}\bar{J}\bar{K}\bar{L}} \nn\\
 \tilde{G}_{\bar{I}\bar{J}\bar{K}11} &=& G_{\bar{I}\bar{J}\bar{K}11}
   -\frac{\sqrt{2}}{16\p}\left(\frac{\k}{4\p}\right)^{2/3}\left(
   \d (x^{11})\o^{(\c ,1)}+\d (x^{11}-\p\r ) \o^{(\c ,2)}
   \right)_{\bar{I}\bar{J}\bar{K}} \label{Gtdef}
\eea
with the gaugino bilinears
\begin{equation}
 \o^{(\c ,i)}_{\bar{I}\bar{J}\bar{K}}=\tr\,\bar{\c}^{(i)}
             \G_{\bar{I}\bar{J}\bar{K}}\c^{(i)}\; .
\end{equation}
In terms of the new field $\tilde{G}$ the relevant part of the
11--dimensional action takes the simple form
\begin{equation}
 S^{(\c )} = -\frac{1}{48\k^2}\int_{M_{11}}\sqrt{-g}\tilde{G}_{IJKL}
             \tilde{G}^{IJKL}\label{ps}
\end{equation}
which reflects the familiar fact that terms containing gaugino
bilinears, together with the original $G^2$--term, group into a perfect
square. In addition, we need the Killing spinor equations in terms of
$\tilde{G}$
\bea
 \d\J_{\bar{I}} &=& D_{\bar{I}}\eta +\frac{\sqrt{2}}{288}\tilde{G}_{JKLM}
                    ({\G_{\bar{I}}}^{JKLM}-8\d_{\bar{I}}^J\G^{KLM})\eta = 0
                    \label{k10}\\
 \d\J_{11} &=& D_{11}\eta +\frac{\sqrt{2}}{288}\tilde{G}_{JKLM}
                    ({\G_{11}}^{JKLM}-8\d_{11}^J\G^{KLM})\eta\nn \\
           && +\frac{1}{192\p}\left(\frac{\k}{4\p}\right)^{2/3} 
              \left[\d (x^{11})\o^{(\c ,1)}+\d (x^{11}-\p\r )
              \o^{(\c ,2)}\right]_{\bar{I}\bar{J}\bar{K}}\G^{\bar{I}
              \bar{J}\bar{K}}\eta = 0
              \label{k11}
\eea
and the equations of motion and the Bianchi identity for $\tilde{G}$
\bea
 D_I\tilde{G}^{IJKL} &=& 0 \label{Geom}\\
 (d\tilde{G})_{11\Ib\Jb\Kb\Lb} &=& 4\sqrt{2}\p\left(\frac{\k}{4\p}
                           \right)^{2/3}\left[J^{(0)}\d (x^{11})+J^{(N+1)}
                           \d (x^{11}-\p\r )+\right.\nn \\
                       &&\left.\qquad\qquad\qquad\qquad\frac{1}{2}
                         \sum_{n=1}^NJ^{(n)}(\d (x^{11}-x_n)+\d (x^{11}+x_n))
                           \right]_{\Ib\Jb\Kb\Lb}\nn\\
                       &&-\frac{\sqrt{2}}{16\p}\left(\frac{\k}{4\p}
                         \right)^{2/3}\left[ J^{(\c ,1)}\d (x^{11})+
                         J^{(\c ,2)}\d (x^{11}-\p\r )\right]_{\bar{I}\bar{J}
                         \bar{K}\bar{L}}\label{GB}
\eea
The sources $J^{(0)}$ and $J^{(N+1)}$ on the orbifold planes are as
usual given by
\begin{equation}
\begin{aligned}
 J^{(0)} &= -\frac{1}{16\p^2}\left.\left(\tr F^{(1)}\wedge F^{(1)} 
      - \frac{1}{2}\tr R\wedge R\right)\right|_{x^{11}=0} \; , \\
 J^{(N+1)} &= -\frac{1}{16\p^2}\left.\left(\tr F^{(2)}\wedge F^{(2)} 
      - \frac{1}{2}\tr R\wedge R\right)\right|_{x^{11}=\p\r} \; .
\end{aligned}
\label{J} 
\end{equation}
The sources $J^{(n)}$, $n=1,\dots ,N$ arise due to the presence of
the five--branes. Finally, the sources $J^{(\c ,i)}$ originate from
the gaugino condensates and are defined as
\begin{equation}
 J^{(\c ,i)} = d\o^{(\c ,i)}\; .
\end{equation}
For generality, we have considered condensates on both orbifold
planes. Later on we will specialize to the ``physical'' case of one
condensate only. Specializing to the field configuration of the above
vacua all sources $J^{(n)}$, where $n=0,\dots ,N+1$, constitute
$(2,2)$--forms on the Calabi--Yau space. Then the charges $\b_i^{(n)}$ that
we have defined earlier can be written as
\begin{equation}
 \b_i^{(n)} = \int_{C_{4i}}J^{(n)}\; .
\label{bzdef}
\end{equation}
where $\{ C_{4i}\}$ is a basis of $H_4(X)$ dual to the basis
$\{C_2^i\}$ of $H_2(X)$ introduced above.

Of course, the non--standard embedding vacua with five--branes,
eq.~\eqref{solform}--\eqref{massless}, that we have constructed 
constitute a solution of the eqs.~\eqref{k10}--\eqref{GB} in the
absence of condensates, as they should. As usual, we will consider
condensates which, as part of the vacuum, are covariantly constant
and closed. Let us now discuss our
vacua in the presence of such condensates. The only change in the
equations (with the replacement of $G$ by $\tilde{G}$ understood) is
then the appearance of the condensate in the Killing spinor
equation~\eqref{k11}. Hence, if we set $\tilde{G}$ equal to the field
$G^{(1)}$ of our solutions and use the corresponding Killing spinor $\eta$ and
metric $g_{IJ}$, we can satisfy all equations except $\d\J_{11}=0$.
Following ref.~\cite{hor}, one can try and modify the spinor $\eta$ to
$\eta +\d\eta$ so as to also satisfy this final equation. One finds from
eq.~\eqref{k11}
\begin{equation}
 \partial_{11}(\d\eta ) = -\frac{1}{192\p}\left(\frac{\k}{4\p}
    \right)^{2/3}\left[ \o^{(\c ,1)}\d (x^{11})+\o^{(\c ,2)}\d(x^{11}-\p\r )
    \right]_{\bar{I}\bar{J}\bar{K}}\G^{\bar{I}\bar{J}\bar{K}}\eta_0\; .
 \label{kd}
\end{equation}
Locally, this equation is solved by $\d\eta$ proportional to
$\o_{\Ib\Jb\Kb}^{(\c ,i)}\G^{\Ib\Jb\Kb}\eta^{({\rm CY})}\e (x^{11})$,
where $\e (x^{11})$ is the step functions which is $+1$ for positive
$x^{11}$ and $-1$ otherwise. Such a modification also leaves the
10--dimensional part of the Killing spinor equation, $\d\J_{\Ib}=0$,
intact since the condensates are covariantly constant. Matching both
$\d$--functions in eq.~\eqref{kd}, however, is impossible unless
\begin{equation}
 (\o^{(\c ,1)}+\o^{(\c ,2)})_{\bar{I}\bar{J}\bar{K}}\G^{\bar{I}\bar{J}\bar{K}}
 \eta_0 = 0\; . \label{unbroken}
\end{equation}
Generally, this condition
is violated and this is particularly so in the ``physical'' situation
of a condensate on one orbifold plane only. Clearly, the obstruction
of solving eq.~\eqref{unbroken} is global in nature and is, of course,
just the ``global'' supersymmetry breaking mechanism discovered by 
Ho\v rava~\cite{hor}. What we have shown here is that this global mechanism
continues to work even for the more general vacua with
non--standard embedding and five--branes. This is not surprising since
the essential part of the argument does not depend on the specific
form of the linearized solution considered, but merely on the fact that
the condensates are covariantly constant and closed. Let us summarize
the essential steps of the argument. We have seen that the five--brane
sources $J^{(n)}$ enter the problem only via the Bianchi
identity~\eqref{GB}. The condensate sources $J^{(\c ,i)}$ in this
Bianchi identity vanish since the condensate is closed. Therefore, the
redefined antisymmetric tensor field $\tilde{G}$ is unchanged from the
previous $G$. As a consequence, the only obstruction to satisfying the
Killing spinor equations comes from the explicit condensate terms in
eq.~\eqref{k11}. Hence, the equation~\eqref{kd} for the
correction to the Killing spinor does not contain the antisymmetric
tensor field $\tilde{G}$, the only object that carries information
about the five--branes. The nature of the breaking is,
therefore, independent on the presence of five--branes.
Although not involving anything essentially new, we found it worth
presenting the above argument in some detail since it
contradicts the ``naive'' expectation. Intuitively, one might
expect that a supersymmetry breaking mechanism global in the
orbifold is modified by five--branes stacked along this direction. What
we have seen is that this does not happen, at least for the pure
background case that we have discussed so far.

\section{The gaugino condensate potential}

In the previous section, we have shown that even for non--standard
embedding and in the presence of five--branes supersymmetry breaking
due to gaugino condensates on the orbifold planes continues to work
via Ho\v rava's global mechanism. Now we would like to proceed further
and compute the low energy superpotential associated with this
breaking. Again, an important question to be clarified is
how the five--branes enter the four--dimensional low energy theory. In
particular, since the five--branes are stacked ``between'' what will
become the hidden and the observable sector at low energy, one might
expect a significant change of the non--perturbative potential. 

To address this problem, let us first recall how the gaugino
condensate superpotential arises in heterotic
M--theory~\cite{gau}. Although the condensate sources $J^{(\c ,i)}$
vanish for the background solution, this is no longer true once the
moduli in this solution are promoted to four--dimensional fields. More
specifically, we use the standard expression~\cite{drsw}
\begin{equation}
 \o^{(\c ,i)}_{abc} = \tr\, \bar{\c}^{(i)}\G_{abc}\c^{(i)} =
                      \L^{(i)}\O_{abc}
 \label{cond} 
\end{equation}
for the condensate, where $\O_{abc}$ is the covariantly constant
$(3,0)$--form of the Calabi--Yau space. In this expression, the
condensation scales $\L^{(i)}$ are functions of the low--energy
moduli fields. Hence, we have the non--vanishing components
\begin{equation}
 J^{(\c ,i)}_{\m abc} = \partial_\m\L^{(i)}\O_{abc}\; \label{condsour}
\end{equation}
of the gaugino sources. In the Bianchi identity~\eqref{GB}, 
these sources appear with $\d$--functions and they have to be
carefully integrated out when deriving the four--dimensional
effective action. As was first observed in ref.~\cite{low,elten}, this
is a general phenomenon for theories with dynamical fields on
boundaries. Not only have boundary sources induced by the background
fields to be taken into account (as we did when we derived the above
vacuum solutions) but, in addition, all sources that appear once the
zero modes are promoted to fields in the low energy theory. Doing this
is essential in order to arrive at the correct low energy
theory and, in particular, to obtain all terms required by
supersymmetry. Applied to the present situation, it means that the
antisymmetric tensor field $\tilde{G}$ should have the form
\begin{equation}
 \tilde{G} = G^{(1)}+G^{({\rm mod})}+G^{(\c )}\; .\label{Gstruct}
\end{equation}
Here $G^{(1)}$ is the background part explicitly given in
eq.~\eqref{sol} and $G^{(\c )}$ originates from integrating out the
condensate sources~\eqref{condsour} in the Bianchi
identity~\eqref{GB}. Explicitly one has
\bea
 G^{(\c )}_{abc11} &=& -\frac{\sqrt{2}}{32\p^2\r}\left(\frac{\k}{4\p}
                   \right)^{2/3}\left[ \L^{(1)}+\L^{(2)}+\l\right]\O_{abc}
 \label{Gc} \\
 G^{(\c )}_{\m abc} &=& -\frac{\sqrt{2}}{32\p}\left(\frac{\k}{4\p}
                        \right)^{2/3}\partial_\m\left[\L^{(1)}-\frac{|x^{11}|}
                        {\p\r}(\L^{(1)}+\L^{(2)})\right]\O_{abc}\; .
 \label{Gc1}
\eea 
Using the flux quantization rule of heterotic M--theory derived in
ref.~\cite{gau}, one can show that the constant $\l$ in this
expression must vanish. Finally, $G^{({\rm mod})}$
accounts for all other sources involving derivatives of low
energy fields as discussed above. It is clear that the gaugino
condensate potential arises by inserting $\tilde{G}$ into the perfect
square~\eqref{ps}. Then, the standard result for the low energy
potential, of course, arises from $G^{(\c )}_{abc11}G^{(\c )abc11}$. One
could, however, expect corrections to this result from mixing terms
between $G^{(\c )}$ and other parts of~\eqref{Gstruct}, in particular
those that originate from the presence of the five--branes or contain
five--brane zero modes. To discuss this in more detail, let us refer to
a part of $\tilde{G}$ with $n$ holomorphic, $m$ antiholomorphic and
$k$ $11$ indices (where $k=0,1$) as a $(n,m,k)$ component. In this terminology,
the standard gaugino potential arises from $(3,0,1)$ and
$(0,3,1)$ components. Let us first consider the background field
$G^{(1)}$. Since the form of $G^{(1)}$ is intimately related to the
presence of the five--branes, a mixing term $G^{(1)}G^{(\c )}$ could
lead to a correction of the potential induced by
five--branes. However, eq.~\eqref{sol} shows that that $G^{(1)}$
consists of $(2,2,0)$, (2,1,1) and (1,2,1) components only and, hence,
such a mixing term is impossible. A part of $G^{({\rm mod})}$ is
related to sources from low energy fields on the orbifold
planes. Explicit expressions for the various components from such
sources have been given in ref.~\cite{low}. There exists, in fact,
one $(3,0,1)$ component proportional to $\O_{abc}W$, where $W$ is the
matter field superpotential. This leads to an expected cross term
between the condensate potential and the matter field superpotential.
In fact, such a term always has to be present and is in no way related
to the presence of five--branes. There is another part of $G^{({\rm
mod})}$ due to sources from low energy fields that originate from the
five--brane worldvolume. The orientation of the five--branes
guarantees, however, that this part of $G^{({\rm mod})}$ is of the
type $(2,2,0)$ and, hence, cannot mix with $G^{(\c )}$. In conclusion,
we find that the gaugino condensate potential does not receive
corrections due to non--standard embedding or the presence of
five--branes. We simply have $W_{\rm gaugino}\sim\L^{(1)}+\L^{(2)}$.

Let us now specialize to the physical case $\L^{(1)}=0$ and
$\L^{(2)}\neq 0$. Following~\cite{drsw}, we write
\begin{equation}
 \L^{(2)}\sim \frac{\a_{\rm GUT}}{\sqrt{v}}\exp\left[ -
               \frac{6\p}{b\a_{\rm GUT}}f^{(2)}\right]
 \label{L2}
\end{equation}
where
\begin{equation}
 \a_{\rm GUT} = \frac{(4\p\k^2)^{2/3}}{2v}\; .
\end{equation}
Here $b$ and $f^{(2)}$ are the $\b$--function coefficient and the
gauge kinetic function for the gauge group on the second orbifold
plane. Hence the superpotential for a single condensate is given by
\begin{equation}
 W_{\rm gaugino} = h\exp\left[ -\frac{6\p}{b\a_{\rm GUT}}f^{(2)}\right]
 \label{Wcond}
\end{equation}
where $h$ is a constant of order $\k /\r^{1/2}v$. For more than one
condensate~\cite{multi}, we get a sum of terms of the above type
with different coefficients $b$, as usual. 

To summarize, let us briefly recapitulate why five--branes did not
effect the potential. Once the zero modes are promoted to low energy
fields, the Bianchi identity~\eqref{GB} contains, in addition to the
five--brane sources, condensate sources as well as other sources
related to the zero modes on the orbifold planes and on the five--brane
worldvolumes. This leads to a complicated structure of $\tilde{G}$ as
indicated by eq.~\eqref{Gstruct}. The conventional part of the gaugino
condensate potential results from $\tilde{G}^2$. Hence, one could
expect cross terms between the gaugino part and parts related to the
presence of five--branes. As we have seen, those cross terms do not
exist because the gaugino part and the five--brane part are associated
to different index structures in the Calabi--Yau space. As a result,
the potential is unmodified.

\section{The five--dimensional viewpoint}

The above line of reasoning can be equivalently presented purely in
the five--dimensional effective action of heterotic
M--theory. Logically, this does not, of course,
provide any new information over what we have done in 11
dimensions. However, since the five--dimensional effective action is of
interest in its own right, we find it useful to reformulate our
argument from a five--dimensional viewpoint.

To do so, we need to recall some facts about the five--dimensional
effective action of heterotic M--theory~\cite{losw,losw1,nse}. The
bulk part of this theory represents a gauged $\cN =1$ supergravity
theory coupled to $h^{1,1}-1$ vector multiplets and $h^{2,1}+1$
hypermultiplets. Five--dimensional space--time has the structure
$M_5=S^1/Z_2\times M_4$ where $M_4$ is a smooth four manifold.

This bulk supergravity is then coupled to two four--dimensional $\cN =1$
theories on the orbifold fix points
that originate from the two $E_8$ sectors of the 11--dimensional
theory and, in addition, to $N$ three--brane worldvolume theories with $\cN =1$
supersymmetry that originate from the five--branes in the
vacuum. For the purpose of this section, we can concentrate on the
structure of the bulk theory. Some details about the spectrum on the
four--dimensional theories will be given in the next section.

To be specific, we introduce the Calabi--Yau $(1,1)$ moduli $a_5^i$ and
associated $(1,1)$ vector fields $\cA^i_\a$, where
$i,j,k,\dots =1,\dots ,h^{1,1}$. We use indices
$\a ,\b ,\g =0,\dots ,3,11$ for the five--dimensional space--time.
In the bulk, the field strengths of these vector fields are given by
\begin{equation}
 \cF^i_{\a\b}=\partial_\a\cA_\b^i-\partial_\b\cA_\a^i\; .
\end{equation}
The volume modulus of the Calabi--Yau space can be expressed in terms of the
$(1,1)$ moduli as
\begin{equation}
 V_5=d_{ijk}a_5^ia_5^ja_5^k/6
\end{equation}
with the triple intersection
numbers $d_{ijk}$. The multiplet structure in five dimensions also
leads us to define the shape moduli
\begin{equation}
 b_5^i=V_5^{-1/3}a_5^i\; .
\end{equation}
Since they satisfy the constraint
\begin{equation}
 \cK (b_5)\equiv d_{ijk}b_5^ib_5^jb_5^k =6\; ,
\end{equation}
they represent only $h^{1,1}-1$ independent degrees of freedom. To make
this more explicit, we introduce independent scalars $\f^x$, where
$x=1,\dots ,h^{1,1}-1$ such that $b_5^i =b_5^i(\f^x)$. With these
definitions, the gravity multiplet has the field content
\begin{equation}
 (g_{\a\b},\frac{2}{3}b_{5i}\cA^i_\a ,\psi^A_\a )\; .
\end{equation}
Here $g_{\a\b}$ is the $D=5$ metric, $\frac{2}{3}b_{5i}\cA^i_\a$ is 
the graviphoton and $\psi^A_\a$ is the gravitino. Fermions in five dimensions
are described by symplectic spinors and, in this section, we use
indices $A,B,C,\dots =1,2$ for the $SU(2)$ R symmetry. Since the
gravity multiplet contains one vector field one remains with $h^{1,1}-1$
vector multiplets. Their structure is given by
\begin{equation}
 (\f^x ,b_{5i}^x\cA_\a^i ,\l^{xA})
\end{equation}
where $b_{5i}^x$ projects onto the
$\f^x$ subspace and $\l^{xA}$ are the gauginos. Note that so far we
have not used the volume modulus $V_5$. It becomes part of the
universal hypermultiplet
\begin{equation}
 (V_5,\s,\x ,\bar{\x},\z^A)\; .
\end{equation}
Here $\s$ is a real scalar and $\z^A$ are the hypermultiplet fermions.
The bulk field strength of the complex scalar field $\x$ is denoted by
\begin{equation}
 X_\a=\partial_\a\x\; .
\end{equation}
This field strength will be of some importance in the following.
We do not need to explicitly introduce the remaining $h^{2,1}$
hypermultiplets.

The gauging of the theory is with respect to the isometry
$\s\rightarrow\s +\mbox{const}$ of the universal hypermultiplet coset space
and the associated gauge boson is a certain linear combination
$\a_i\cA^i_\a$ of the vectors fields, where $\a_i$ are charges. The
bulk action for those fields has been derived in
ref.~\cite{losw,losw1} for the case of the standard embedding. It has
been shown~\cite{nse} that this action remains valid even in the
presence of five--branes once the charges $\a_i$ are chosen
appropriately. More specifically, for  the interval
$z_n\leq |z|\leq z_{n+1}$ between any two neighboring five--branes,
where $n=0,\dots ,N$, one should replace the charges $\a_i$ by
\begin{equation}
 \a_i\rightarrow\a_i^{(n)}\equiv\frac{\sqrt{2}{\e }}{\r}\sum_{m=0}^n\b_i^{(n)}
 \label{repl}
\end{equation}
with $\b_i^{(n)}$ as introduced in eq.~\eqref{instanton}. Strictly
speaking, for each interval one has, therefore, a different gauged
supergravity theory characterized by the appropriate set of gauge charges.

Now, we need to set up the essential parts of the $D=5$ theory
incorporating a gaugino condensate. As in 11 dimensions, we start with
a field redefinition to avoid singularities. Since the condensate
originates from the $(3,0)$ part of the Calabi--Yau space, as in
eq.~\eqref{cond}, we need to redefine the corresponding zero mode. This
zero mode is precisely the complex scalar field $\x$  with field
strength $X_\a$ in the universal hypermultiplet. In analogy with
eq.~\eqref{Gtdef}, we define
\bea
 \tilde{X}_\m &=& X_\m \\
 \tilde{X}_{11} &=& X_{11}+\frac{\sqrt{2}}{32\p}\frac{\k_5^2}
                    {\a_{\rm GUT}}\left[\l^{(1)}\d (x^{11})+\l^{(2)}
                    \d (x^{11}-\p\r )\right]\label{Xt}
\eea
where $\k_5^2=\k^2/v$ is the five--dimensional Newton constant and
$\a_{\rm GUT}=(4\p\k^2 )^{2/3}/2v$. Then the analog in five dimensions
of the perfect square~\eqref{ps} reads
\begin{equation}
 S^{(\c )}_5 = -\frac{1}{\k_5^2}\int_{M_5}\sqrt{-g}\, V^{-1}
               \tilde{X}_\a\tilde{X}^\a\; .\label{psX}
\end{equation}
The Bianchi identity for $X_\a$ has been given in ref.~\cite{losw1}.
Inserting the definition~\eqref{Xt} we obtain
\begin{equation}
 \left(d\tilde{X}\right)_{11\m} = -\frac{\k_5^2}{16\sqrt{2}\p
    \a_{\rm GUT}}\left[(4J_\m^{(1)}+\partial_\m\L^{(1)})\d (x^{11})
    +(4J_\m^{(2)}+\partial_\m\L^{(2)})\d (x^{11}-\p\r )\right]
    \label{XBI}
\end{equation}
where $J_\m^{(i)}$ is proportional to $\partial_\m W^{(i)}$, the
derivative of the matter field superpotential on the $i$--th orbifold
plane. It is important to note that neither the Bianchi
identity~\eqref{XBI} nor the relevant part of the equation of motion
for $\tilde{X}_\a$ (to be derived from \eqref{psX}) have sources
located on the three--branes that originate from the five--branes. The
reason for this was already explained from the 11--dimensional
perspective. Due to the orientation of the five--branes, they
cannot contribute to the $(3,0)$ part of the Calabi--Yau space. In the
five--dimensional theory, this fact is manifest in the structure of the
above equations.

Finally, we need the five--dimensional bulk supersymmetry
transformations. For the standard embedding, they have been derived in
ref.~\cite{losw,losw1}. To adapt these transformations to our
situation, they should be modified in two ways. First, one should apply
the replacement~\eqref{repl} as discussed. Secondly, one should
incorporate the gaugino condensate and the definition~\eqref{Xt}, as in
the analogous 11--dimensional equations~\eqref{k10} and \eqref{k11}.
This leads to
\bea
\d \psi_\a^A &=& \nabla_\a\e^A 
     + \frac{\sqrt{2}i}{8}
          \left({\g_\a}^{\b\g}-4\d_\a^\b\g^\g\right)b_{5i}\cF_{\b\g}^i\e^A
     - ({{P_\a}^A}_B+{{S_\a}^A}_B) \e^B \nn \\ && \qquad
     - \frac{\sqrt{2}}{12}V^{-1}b_5^i\a_i^{(n)}\g_\a\,
       {{\t_3}^A}_B\e^B \\
\d \l^{xA} &=& b_{5i}^x\left(-\frac{1}{2}i\g^\a\partial_\a b_5^i\e^A
             -\frac{1}{2\sqrt{2}}\g^{\a\b}\cF_{\a\b}^i\e^A
             -\frac{i}{2\sqrt{2}}V_5^{-1}\a^{(n)i}{{\t_3}^A}_B\e^B
             \right) \\
\d \z^A &=& 
     - i {{Q_\a}^A}_B \g^\a \e^B
     - \frac{i}{\sqrt{2}}b_5^i\a_i^{(n)}V_5^{-1}{{\t_3}^A}_B\e^B
\label{susy5}
\eea
in the interval $z_n\leq |z|\leq z_{n+1}$, where $n=0,\dots ,N$.
Here $\t_i$ with  $i=1,2,3$ are the Pauli spin matrices and we have
defined
\begin{equation}
\begin{aligned}
   {{P_\a}^A}_B &= {\left( \begin{array}{cc} 
           \frac{\sqrt{2}i}{96}V_5 \e_{\a\b\g\d\e}G^{\b\g\d\e} & 
           V_5^{-1/2}\tilde{X}_\a \\
           - V_5^{-1/2}\bar{\tilde{X}}_\a & 
           - \frac{\sqrt{2}i}{96}V_5 \e_{\a\b\g\d\e}G^{\b\g\d\e}
        \end{array} \right)^A}_B \\
   {{Q_\a}^A}_B &= {\left( \begin{array}{cc}
           \frac{\sqrt{2}i}{48}V_5 \e_{\a\b\g\d\e}G^{\b\g\d\e} 
               - \frac{1}{2}V_5^{-1}\pt_\a V_5 &
           V_5^{-1/2}\tilde{X}_\a \\
           V_5^{-1/2}\bar{\tilde{X}}_\a & 
           \frac{\sqrt{2}i}{48}V_5 \e_{\a\b\g\d\e}G^{\b\g\d\e} 
               + \frac{1}{2}V_5^{-1}\pt_\a V_5 
        \end{array} \right)^A}_B\; .
\end{aligned}
\end{equation}
The contribution from the condensate is encoded in ${{S_\a}^A}_B$ with
the non--zero components
\begin{equation}
 {(S_{11})^1}_2=-{(\bar{S}_{11})^2}_1=-\frac{3\k_5^2}{32\sqrt{2}\p
                 \a_{\rm GUT}}V_5^{-1/2}\left[\L^{(1)}\d (x^{11})+
                 \L^{(2)}\d (x^{11}-\p\r )\right]\; .
\end{equation}
This concludes our review of the five--dimensional effective action.
Let us now discuss the vacuum solution of this theory in the absence
of a condensate. This solution is a BPS
multi--domain wall with $N+2$ worldvolumes that constitutes the
appropriate background for a reduction to four dimensions. It is a
straightforward generalization of the solution found in
ref.~\cite{losw,losw1} and it can be most easily obtained, in its
linearized version, by just rewriting the 11--dimensional
solution~\eqref{solform}--\eqref{massless}
in terms of five--dimensional fields. Moreover, it satisfies the
Killing spinor equations $\d \psi_\a^A=0$, $\d \l^{xA}=0$ and
$\d \z^A=0$ for a certain Killing spinor $\e^A$ as long as the
condensate vanishes. Once the condensate is switched on, the only
supersymmetry transformation that changes is the one for
$\psi_{11}^A$. As in 11 dimensions, we can try to modify the
spinor $\e^A$ to $\e^A+\d\e^A$ so as to compensate for this change.
This leads to
\begin{equation}
 \partial_{11}(\d\e^{1}) = {(S_{11})^1}_2\,\e^{2}\; .
\end{equation}
and the equation for $\d\e^{2}$ obtained by conjugation.
Again, this equation has a local solution everywhere. The existence of
a global solution, however, requires
\begin{equation}
 \L^{(1)}+\L^{(2)} = 0
\end{equation}
which is the five--dimensional analog of eq.~\eqref{unbroken}. Hence,
we have rediscovered the global breaking mechanism from a
five--dimensional viewpoint.

Now, we promote the moduli of the multi--domain wall solution to
four--dimensional fields. Then, the sources in the Bianchi
identity~\eqref{XBI} become non--vanishing. The solution for
$\tilde{X}_\a$ then takes the form
\begin{equation}
 \tilde{X}_\a = \tilde{X}_\a^{(W)}+\tilde{X}_\a^{(\c )}
\end{equation}
where $\tilde{X}_\a^{(W)}$ accounts for the sources $J_\m^{(i)}$
related to the superpotentials on the orbifold planes and
$\tilde{X}_\a^{(\c )}$ originates from the condensate sources.
The solution for $\tilde{X}_\a^{(\c )}$ is given by
\bea
 \tilde{X}_{11}^{(\c )} &=& -\frac{\k_5^2}{32\sqrt{2}\p^2\r
     \a_{\rm GUT}}\left(\L^{(1)}+\L^{(2)}\right) \\
 \tilde{X}_\m^{(\c )} &=& -\frac{\k_5^2}{32\sqrt{2}\p
     \a_{\rm GUT}}\partial_\m\left(\L^{(1)}-\frac{|x^{11}|}{\p\r}
     (\L^{(1)}+\L^{(2)})\right)\; .
\eea
Then, the gaugino condensate potential arises from inserting
$\tilde{X}_{11}^{(\c )}$ into the perfect square~\eqref{psX}. We find
$W_{\rm gaugino}\sim \L^{(1)}+\L^{(2)}$ as before. Hence, for the
``physical'' case $\L^{(1)}=0$ and $\L^{(2)}\neq 0$, we obtain
the previous result~\eqref{Wcond} for the superpotential in terms of
the gauge kinetic function. Once more
we see that the potential is unchanged, at this level, by
the three--branes that originate from the five--branes.
From the five--dimensional viewpoint, the reason is that these
three--branes decouple from the crucial Bianchi identity~\eqref{XBI}
and the equation of motion for $\tilde{X}_\a$, as we have stated earlier.

\section{The four--dimensional effective action}
 
To proceed further, we need some information about the
four--dimensional $\cN =1$ effective action associated with our vacua.
In particular, we need to know the gauge kinetic function that enters
the gaugino condensate potential~\eqref{Wcond}. The main goal of this
section is, therefore, to derive the matter field part of the low energy
action. To put these results into the correct context we will,
however, start by explaining the general structure of the low energy
actions derived from our vacua~\cite{nse}.

\vspace{0.4cm}

The field content of the action splits into $N+2$ sectors coupled to
one another only gravitationally. Two of those sectors arise
from the $E_8$ gauge multiplets on the orbifold planes while the other
$N$ originate from the degrees of freedom on the five--brane
worldvolume theories. All of these sectors have $\cN =1$ supersymmetry.
Let us first explain the two ``conventional''
$E_8$ sectors. We have considered internal gauge bundles $V_i$,
$i=1,2$ with gauge groups $G_i$ in these two sectors. The surviving
low energy gauge groups $H_i$ are the commutants of $G_i$ in
$E_8$. In general, from each boundary $i$, we have gauge matter fields
in those representations $R$ of $H_i$ that appear in the decomposition
$\mbf{248}_{E_8}\rightarrow\sum_{S,R}(S,R)$ of the adjoint of $E_8$
under $G_i\times H_i$. The number of families, for a certain
representation $R$, is given by the dimension of $H^1(X,V_{iS})$ where
$V_{iS}$ is the vector bundle $V_i$ in the representation $S$. 
We stress again that, for the class of vacua we are considering here,
we expect many examples where the chiral asymmetry takes the ``physical''
value three.

From the five--branes, we get additional gauge groups $\cH^{(n)}$,
$n=1,\dots N$. Generically, these groups are given by
$\cH^{(n)}=U(1)^{g_n}$ where $g_n$ is the genus of the cycle
$C^{(n)}\in X$ on which the $n$--th five--brane wraps. They can,
however, enhance to non--Abelian groups, typically unitary groups,
when five--branes overlap or a curve $C^{(n)}$ degenerates. In any
case, the low energy gauge group is enlarged to $H = H_1\times
H_2\times \cH^{(1)}\times\dots\times\cH^{(N)}$. 

Various sets of moduli arise in the low energy theory. First of all, we
have the K\"ahler and complex structure moduli of the Calabi--Yau
space along with their superpartners. From the gauge bundles $V_i$ we
get two sets of bundle moduli. Furthermore, we have moduli describing
the position of the five--branes in the orbifold direction as well as
moduli that parameterize the moduli space of the five--brane curve
within the Calabi--Yau space.

It is quite possible that reasonable observable sectors could arise
from the five--brane worldvolume, although explicit models along these
lines have not yet been explored. In this paper, we adopt the more
``conservative'' viewpoint that one of the $E_8$ sectors, say the one
on the first orbifold plane, constitutes the observable sector while
all other sectors are hidden. We have shown~\cite{np} that low energy
grand unification groups (for example $H_1 = SU(5)$) with three
generations can easily be obtained in this sector due to the freedom
introduced by the presence of five--branes. We would then like to
study the low energy (matter field) action in this sector and
supersymmetry breaking induced by the other $E_8$ sector. It is likely
that the hidden sectors from five--branes can contribute to
supersymmetry breaking via gaugino condensation as well under certain
circumstances. We postpone the investigation of this question to a future
publication~\cite{inpr}. 

\vspace{0.4cm}

Let us now be more specific about the ingredients we need to discuss
the effective theory. For the observable gauge matter, we focus on a
specific irreducible representation $R$ of the gauge
group $H_1$. Suppose that $R$ appears as the product
$S\times R$ in the decomposition of $\mbf{248}_{E_8}$ under
$G_1\times H_1$. Then there will be
$\mbox{dim}(H^1(X,V_{1S}))$ families of this type which we denote by
$C^{Ip}(R)$. Here $I,J,K,\dots = 1,\dots ,\mbox{dim}(H^1(X,V_{1S}))$
is the family index and $p,q,\dots =1,\dots ,\mbox{dim}(R)$ the
representation index of $R$. We also need to introduce a basis
$\{ u_I^x(R)\}$ of $H^1(X,V_{1S})$, where $x,y,\dots =1,\dots ,\mbox{dim}(S)$
is the representation index of $S$, and generators $T_{xp}(R)$
corresponding to the $(R,S)$--part in the decomposition of
$\mbf{248}_{E_8}$. The complex conjugate of these generators is
denoted by $T^{xp}(R)$ and the normalization is chosen such that
$\mbox{tr}(T_{xp}(R)T^{yq}(R))=\d_x^y\d_p^q$. As an example, consider
a gauge bundle $V_1$ with gauge group $G_1=SU(4)$. Then we have a low energy
theory with the grand unification group $H_1=SO(10)$. The
corresponding decomposition of the adjoint of $E_8$ under
$SU(4)\times SO(10)$ reads
\begin{equation}
\mbf{248} = (\mbf{15},\mbf{1}) \oplus (\mbf{1},\mbf{45}) \oplus
            (\mbf{6},\mbf{10}) \oplus
           (\mbf{4},\mbf{16}) \oplus (\mbf{\bar{4}},\mbf{\bar{16}})\; .
\end{equation}
The relevant representations are then
$(S,R)=(\mbf{4},\mbf{16})$ for the families,
$(S,R)=(\mbf{\bar{4}},\mbf{\bar{16}})$ for the anti--families and
$(S,R)=(\mbf{6},\mbf{10})$ for the Higgs fields.

Also, we need to introduce
explicitly some of the moduli. Most notably, there are the averaged
Calabi--Yau volume $V$  and the radius $R$ of the orbifold in units of
$v$ and $\r$, respectively. The $(1,1)$--moduli $a^i$ of the
Calabi--Yau space, where $i,j,k,\dots =1,\dots ,h^{1,1}$, are defined
as $\o =a^i\o_i$. Here $\o_{a\bbar}=-ig_{a\bbar}^{({\rm CY})}$ is the
K\"ahler form. As a function of the $(1,1)$--moduli, the volume is
given by $V=V(a)=\cK (a)/6$ with the K\"ahler potential
$\cK (a)=d_{ijk}a^ia^ja^k$ and the Calabi--Yau intersection numbers
$d_{ijk}$. We also note that the metric~\eqref{metric11} can be
expressed as
\begin{equation}
 G_{ij}(a) = -\frac{1}{2}\partial_i\partial_j\ln\cK (a)\; .
\end{equation}
An important role will be played by the moduli $z^n\in [0,1]$, where
$n=1,\dots N$, that specify the position of the five--branes in the
orbifold direction. 

Introducing the conventional four--dimensional chiral fields $S$,
$T^i$ as well as the chiral fields $Z^n$, we have
\begin{equation}
 \RR (S) =V\; ,\qquad \RR (T^i) = Ra^i\; ,\qquad \RR (Z^n)=z^n\; .
\end{equation}
Another relevant quantity is the metric
\begin{equation}
 G_{IJ}^{(R)}(a) = \frac{1}{vV}\int_X\sqrt{g^{({\rm CY})}}g^{({\rm
             CY})a\bbar}u_{Iax}(R)u_{J\bbar}^x(R) \label{bmetric}
\end{equation}
on the moduli space $H^1(X,V_{1S})$.
In general, this metric is a function of the K\"ahler moduli
$a^i$, as we have indicated, as well as a function of the complex
structure and bundle moduli. Finally, we need the Yukawa--couplings
\begin{equation}
 \l_{IJK}^{(R_1R_2R_3)} = \int_X\O\wedge u_I^x(R_1)\wedge u_J^y(R_2)
                         \wedge u_K^z(R_3)f_{xyz}^{(R_1R_2R_3)}
\end{equation}
where $f_{xyz}^{(R_1R_2R_3)}$ projects onto the singlet in
$R_1\times R_2\times R_3$ (if any). As an example, for the above
$SO(10)$ case, the relevant products are of course
$\mbf{10}\times\mbf{16}\times\mbf{16}$ and
$\mbf{10}\times\mbf{\bar{16}}\times\mbf{\bar{16}}$.
As is well known, these Yukawa--couplings are quasi--topological,
depending on the complex structure moduli and the Dolbeault cohomology
classes of the $u_I^x(R)$ only.

\vspace{0.4cm}

For the explicit computation, it is useful to write the
metric of our vacuum solution in terms of the moduli introduced above.
One finds from eq.~\eqref{solform}--\eqref{bser} 
\bea
 g_{\m\n} &=& V^{-1}R^{-1}\left( 1+\frac{2\sqrt{2}}{3}b_ia^i\right)
              g_{\m\n}^{(4)} \\
 g_{a\bbar} &=&i\left(a^i-\frac{4\sqrt{2}}{3}b_ja^ja^i+\sqrt{2}b^i\right)
               \o_{ia\bbar}
\eea
where the $b_i$ are the parameters encoding the deformation of the
Calabi--Yau space as defined in eq.~\eqref{massless} and
$g_{\m\n}^{(4)}$ is the four--dimensional Einstein--frame metric.
The part of the $E_8$ gauge field strength that gives rise to the
matter fields can be written as
\begin{equation}
 F_{\m\bbar} = \sqrt{2\p\a_{\rm GUT}}\, \sum_R u^x_{I\bbar}(R)\, T_{xp}(R)
               (D_\m C(R))^{Ip}\; .
\end{equation}
Using these expressions in the 10--dimensional Yang--Mills action
\begin{equation}
 S_{\rm YM} = -\frac{1}{8\p\k^2}\left(\frac{\k}{4\p}\right)^{2/3}
              \sum_{i=1}^2\int_{M_{10}^{(i)}}\sqrt{-g}\,\mbox{tr}{F^{(i)}}^2
\end{equation}
one can compute the relevant low--energy quantities. In the following,
we suppress the representation $R$ as well as the
corresponding group indices in order to simplify the
notation. Instead, one can think of the indices $I,J,K\dots$ as
running over the various relevant representations as well as over the
families. We find for the gauge kinetic function $f^{(1)}$ of the
observable group $H^{(1)}$ and the gauge kinetic function $f^{(2)}$ of
the hidden group $H^{(2)}$~\cite{nse}
\bea
 f^{(1)} &=& S+\e T^i\left(\b^{(0)}_i+\sum_{n=1}^{N}(1-Z^n)^2
             \b_i^{(n)}\right) \label{f1}\\
 f^{(2)} &=& S+\e T^i\left(\b^{(N+1)}_i+\sum_{n=1}^{N}(Z^n)^2
             \b_i^{(n)}\right) \label{f2}
\eea
The K\"ahler potential for the observable matter
fields has the structure 
\bea
 K_{\rm matter} &=& Z_{IJ}\bar{C}^IC^J \\
 Z_{IJ} &=& e^{-K_T/3}\left[
            K_{BIJ}-\frac{\e }{S+\Sb}\tilde{\G}_{BIJ}^i
            \sum_{n=0}^{N+1}(1-z^n)^2\b^{(n)}_i\right]\label{K}
\eea
with the $(1,1)$ K\"ahler potential and K\"ahler metric
\bea
 K_T &=& -\ln\left(\frac{1}{6}d_{ijk}(T^i+\Tb^i)(T^j+\Tb^j)
       (T^k+\Tb^k)\right) \\
 K_{Tij} &=& \frac{\partial^2K_T}{\partial T^i\partial\Tb^j}
\eea
and the following quantities associated with the vector bundle
\bea
 K_{BIJ} &=& G_{IJ}(T+\Tb) \\
 \G_{BIJ}^i &=&K_T^{ij}\frac{\partial K_{BIJ}}{\partial T^j} \label{Kdef}\\
 \tilde{\G}_{BIJ}^i &=& \G_{BIJ}^i-(T^i+\Tb^i)Z_{BIJ}-\frac{2}{3} 
                       (T^i+\Tb^i)(T^k+\Tb^k)K_{Tkj}\G_{BIJ}^j\; .
\eea
Note that the metric $K_{BIJ}$ equals the bundle
metric~\eqref{bmetric} viewed as a function of $T^i+\bar{T}^i$.
Finally, for the superpotential of the observable matter, we have the
usual expression
\begin{equation}
 W_{\rm matter} = \frac{1}{3}\tilde{Y}_{IJK}C^IC^JC^K
\end{equation}
where
\begin{equation}
 \tilde{Y}_{IJK} = 2\sqrt{2\p\a_{\rm GUT}}\,\l_{IJK}\; .\label{WC}
\end{equation}
We remark that the K\"ahler potential for hidden matter can easily be
obtained from eq.~\eqref{K} by the replacement $z^n\rightarrow 1-z^n$.

We can now insert the result~\eqref{f2} for the hidden sector gauge
kinetic function into eq.~\eqref{Wcond}. This leads to the explicit
gaugino condensate potential
\begin{equation}
 W_{\rm gaugino} = h\exp\left[ -\frac{6\p}{b\a_{\rm GUT}}\left(
               S+\e T^i\sum_{n=0}^{N+1}(Z^n)^2\b_i^{(n)}\right)
               \right]
 \label{Wcondex}
\end{equation}
for a single condensate. Although the presence of five--branes did
thus far not effect the condensate potential, we see that this final step
introduces a five--brane dependent modification resulting from the threshold
correction to the gauge kinetic function. As a result, the
potential~\eqref{Wcondex} not only depends on the dilaton $S$ and the
$T$--moduli, but also on the five--brane position moduli $Z^n$.

\section{Phenomenological issues} 

In this section we discuss some phenomenological issues related to
supersymmetry breaking and the effective four--dimensional matter
field action that we have computed. Our emphasis will be on the
consequences for soft supersymmetry breaking terms. 

We will not attempt to construct explicit gaugino condensate models
based on the superpotential~\eqref{Wcondex}, since that would lead us
into complicated model building. Instead, we will parameterize
supersymmetry breaking in terms of the auxiliary fields of the moduli.
Usually, this is done using the auxiliary fields associated to the
dilaton and the $T$--moduli~\cite{kl,ST}. The presence of the
five--brane position
moduli $Z^n$ in the gaugino condensate superpotential, however,
suggests that one should, in general, also allow their auxiliary
components to contribute to supersymmetry breaking. Hence, we will
analyze supersymmetry breaking that can be parameterized by
auxiliary fields $(F^S,F^i,F^n)$ corresponding to the moduli
$(S,T^i,Z^n)$, respectively. 

The starting point is the K\"ahler potential
\begin{equation}
 K=\k_P^{-2}K_{\rm mod}+Z_{IJ}\bar{C}^IC^J
\end{equation}
where $K_{\rm mod}$ is the K\"ahler
potential of the moduli depending on $S$, $T^i$, $Z^n$ and the
other moduli and $Z_{IJ}$ was given in eq.~\eqref{K}.
The four--dimensional Newton constant $\k_P$ is defined in terms of
11--dimensional quantities as $\k_P^2=\k^2/2\p\r v$.
One also needs the gauge kinetic function $f^{(1)}$, eq.~\eqref{f1} and the
Yukawa couplings $\tilde{Y}_{IJK}$ in the superpotential~\eqref{WC}.
After integrating out the moduli, the effective Yukawa coupling will be
\begin{equation}
 Y_{IJK} = e^{K_{\rm mod}/2}\tilde{Y}_{IJK}\; .
\end{equation}
It will be useful for our discussion to explicitly list the general
expressions for the soft masses~\cite{sw,gm,kl}. One has
\bea
 m_{3/2}^2 &=& \frac{1}{3}K_{{\rm mod},a\bbar} 
             F^a\Fb^{\bbar} \label{m32gen}\\
 m_{1/2} &=& \frac{1}{2\mbox{Re}(f^{(1)})}F^a\partial_af^{(1)} 
             \label{m12gen}\\
 m^2_{I\bar{J}} &=& m_{3/2}^2Z_{I\bar{J}}-F^a\Fb^{\bbar}
                    R_{a\bbar I\bar{J}} \label{m0gen}\\
 A_{IJK} &=& F^a\left( \partial_aY_{IJK}+\frac{1}{2}K_{{\rm mod},a}Y_{IJK}
                -\G_{a(I}^NY_{JK)N}\right) \label{Agen}
\eea
where
\bea
 R_{a\bbar I\bar{J}} &=& \partial_a\bar{\partial}_{\bbar} Z_{I\bar{J}}
                         -\G_{aI}^NZ_{N\bar{L}}
                         \bar{\G}_{\bbar\bar{J}}^{\bar{L}} \\
 \G_{aI}^N &=& Z^{N\bar{J}}\partial_aZ_{\bar{J}I} \; .
\eea
Here $m_{3/2}$ is the gravitino mass, $m_{1/2}$ is the gaugino mass,
$m_{I\bar{J}}^2$ are the scalar soft masses and $A_{IJK}$ are the
trilinear soft couplings. Indices $a,b,c,\dots$ run over all moduli fields.

There are two aspects of our scenario that make an analysis of soft
terms interesting; that is, the possibility of supersymmetry breaking in
the direction of the five--brane position moduli and the presence of
the order $\e $ correction to the effective action. Both aspects
might influence the magnitude of the soft terms compared to
perturbative heterotic string, as we will discuss later on in more
detail. For standard--embedding vacua of heterotic M--theory and a
simple model with one family, soft terms in the presence of
$\e $ corrections have been already studied~\cite{hp,ckm,gau}.

Before we come to that, we would like to address the problem of
universality of soft breaking parameters. In fact, universality might
provide a stronger phenomenological constraint than just the overall size
of soft terms. We remark that for a
perfectly universal model (at the unification scale) we should have
$m^2_{I\bar{J}}\sim Z_{I\bar{J}}$ and $A_{IJK}\sim Y_{IJK}$. The
obstruction to universality results from the last term in the
expressions for $m^2_{I\bar{J}}$ and $A_{IJK}$ in
eq.~\eqref{m0gen} and \eqref{Agen}. Of course, there is nothing wrong
with choosing the moduli so as to make these last terms universal as
well (something one expects to be able to do for a sufficiently large
number of moduli). However, given the accuracy at which
universality holds, this requires a significant fine tuning in moduli
space. Therefore, it is much more desirable to have a reason for
universality to occur in a given model. The usual assumption along
these lines is that supersymmetry is broken in the direction of moduli
that are flavor--blind. To discuss this possibility, let us neglect the
$\e $ corrections to our effective action for the moment and
concentrate on the lowest order part. Clearly, the lowest order part of
the matter field K\"ahler potential is independent on $S$ and
$Z^n$. Therefore, if supersymmetry breaking occurs in the dilaton
direction the soft terms will be universal to that order~\cite{kl}.
The same is true if some of the auxiliary fields $F^n$ associated
to the five--brane positions are non--vanishing. This last option
represents a new breaking pattern that is directly related to the
presence of five--branes in our vacua. What is the reason for this
structure from an 11--dimensional viewpoint? Both, the $T$--moduli
and the matter fields arise as zero modes associated to the shape
of the Calabi--Yau space. Hence supersymmetry breaking in the $T^i$
directions is generally not flavor--blind. On the other hand, the
moduli $Z^n$ arise as zero--modes of the five--brane worldvolume
theories and, hence, do not distinguish between the generations to
zeroth order.

Let us now discuss the effect of the $\e $ corrections, which we have
neglected so far. Of course, if $\e $ is very small these terms need
not concern us much. However, it is well known that $\e $ is of order
one at the ``physical'' point where coupling unification can be
realized~\cite{bd}. Hence, those terms are quite possibly subject
to the universality constraint as well. Unfortunately, from
eq.~\eqref{K} we see that the $\e $ corrections to the K\"ahler metric
depend on $S$ as well as on $Z^n$. In addition, the zero and first
order contributions to $Z_{IJ}$ are generically not proportional to
each other. This implies that, unlike for the lowest order case,
there is no specific pattern of supersymmetry breaking that will lead
to universality, in general. Note, that also the $F^S$ scenario will
not work in this case. Let us again discuss this from an
11--dimensional point of view. To zeroth order, moduli were
flavor--sensitive or flavor--blind depending on whether or not they
arise as zero modes of the Calabi--Yau space. This does not remain true
once the $\e $ corrections are taken into account. These corrections
correspond to an orbifold dependence of the 11--dimensional background
that arises from the sources on the orbifold planes and the
five--branes. This dependence encodes ``interaction'' across the
orbifold and leads, for example, to a coupling of the five--brane
moduli to the matter fields. This coupling appears despite the fact
that the matter fields and the five--brane moduli are separated in
the orbifold direction. We remark that one expects such corrections
quite generally in vacua that contain branes although they need not to
be necessarily large.

One way out of the universality problem is to find a way to make the zeroth and
first order part of $Z_{IJ}$ proportional. There is,
indeed, a simple way to do that, namely to choose a Calabi--Yau space
with one K\"ahler modulus only ($h^{1,1}=1$). Such Calabi--Yau spaces
exist, the quintic hypersurface in $CP^4$ being the simplest
example. Moreover, as our previous experience shows~\cite{np}, with the
freedom introduced by allowing non--standard embedding and
five--branes the construction of three--generation models on such
Calabi--Yau spaces might be considerably easier than for the
restrictive standard--embedding. Let us, therefore, analyze this
case in more detail. Since we have only one $T$--modulus, the metric
$K_{BIJ}$ in eq.~\eqref{Kdef} can be written as
\begin{equation}
 K_{BIJ}=\frac{3}{(T+\Tb )^2}H_{IJ}
\end{equation}
with some $T$--independent metric $H_{IJ}$. We then find for the
matter field K\"ahler metric
\begin{equation}
 Z_{IJ} =\left[\frac{3}{T+\Tb}+\frac{\e \z}{S+\Sb}\right]
         H_{IJ}
\end{equation}
where
\begin{equation}
 \z=\b^{(0)}+\sum_{n=1}^N(1-z^n)^2\b^{(n)}\; .\label{h}
\end{equation}
The relevant part of the moduli K\"ahler potential then reads
\begin{equation}
 K_{\rm mod} = -\ln (S+\Sb)-3\ln (T+\Tb)+K_5\label{K5}
\end{equation}
where $K_5$ is the K\"ahler potential for the five--brane moduli $Z^n$.
Now $S$, $T$ as well as the five--brane moduli $Z^n$ enter the
matter K\"ahler potential in a generation--independent way and we expect
universal soft masses as long as supersymmetry is broken in the sector
$(F^S,F^T,F^n)$ (which was our original assumption). More specifically
we get
\begin{equation}
 m_{3/2}^2 = \frac{|F^S|^2}{3(S+\Sb)^2}+\frac{|F^T|^2}
             {(T+\Tb)^2}+\frac{1}{3}K_{5n\mb}F^n\Fb^{\mb}
\end{equation}
for the gravitino mass. The gaugino mass is given by
\begin{equation}
 m_{1/2} = m_{1/2}^{(0)}+\e m_{1/2}^{(1)}
\end{equation}
with
\bea
 m_{1/2}^{(0)} &=& \frac{F^S}{2(S+\Sb)} \\
 m_{1/2}^{(1)} &=& \frac{\z F^T}{2(S+\Sb)}-\frac{(T+\bar{T})\z F^S}{2(S+\Sb)^2}
                   +\frac{T\z_nF^n}{2(S+\Sb)}\; .
\eea
For the scalar soft masses we get
\begin{equation}
 m_{IJ}^2 = ({m^{(0)}}^2+\e {m^{(1)}}^2)H_{IJ}
\end{equation}
with
\bea
 {m^{(0)}}^2 &=& \frac{|F^S|^2}{(S+\Sb)^2(T+\Tb)}
                 +\frac{K_{5n\mb}F^n\Fb^{\mb}}{T+\Tb}\\
 {m^{(1)}}^2 &=& -\frac{5\z |F^S|^2}{3(S+\Sb )^3}+
                 \frac{2\z\mbox{Re}(F^S\Fb^{\Tb})}{(S+\Sb )^2(T+\Tb )}
                 +\frac{\z K_{5n\mb}F^n\Fb^{\mb}}{3(S+\Sb )}\nn \\
             &&-\frac{\z_{n\mb}F^n\Fb^{\mb}}{S+\Sb}
               +\frac{2\mbox{Re}(F^S\Fb^{\nb}\z_{\nb})}{(S+\Sb)^2}
               -\frac{2\mbox{Re}(F^T\Fb^{\nb}\z_{\nb})}
               {(T+\Tb )(S+\Sb )}\; .
\eea
Finally, for the trilinear soft terms we find
\begin{equation}
 A_{IJK} = (A^{(0)}+\e A^{(1)})Y_{IJK}
\end{equation}
where
\bea
 A^{(0)} &=& -\frac{F^S}{S+\Sb}+F^nK_{5n} \\
 A^{(1)} &=& -\frac{\z F^T}{S+\Sb}+\frac{(T+\Tb )\z F^S}{(S+\Sb )^2}
             -\frac{(T+\Tb )\z_nF^n}{S+\Sb}\; .
\eea
In these equations, $\z$ contains the five--brane moduli $Z^n$ and
is given by eq.~\eqref{h}. We also used an index notation for
derivatives with respect to $Z^n$, for example
$\z_n=\partial \z/\partial Z^n$.
We see that, by means of a simple topological assumption and
supersymmetry breaking in the $(F^S,F^T,F^n)$ sector, we have achieved
universality to linear order in $\e $ (that is, to the order we can
calculate). Depending on the specific breaking pattern, the size of the
various soft terms is either of order $m_{3/2}$ or of order
$\e m_{3/2}$. Given that $\e $ still has to be somewhat smaller than
one in order to stay within the limits of our approximations, the
second option implies a certain suppression. Let us discuss some
examples. Consider supersymmetry breaking in the $F^T$ direction. Then
the gaugino mass is of order $\e m_{3/2}$, an observation that was
first made in ref.~\cite{hp}. Also, the trilinear soft terms are of
that order while the scalar soft masses still vanish~\cite{ckm,gau}.
As an alternative possibility, consider supersymmetry breaking in the
directions $F^n$ of the five--brane moduli. The gaugino masses then
remain at the order $\e m_{3/2}$ while trilinear soft terms and
scalar masses are now of order $m_{3/2}$. In general, we see that the
structure of our models provides a direct correlation between the magnitude of
soft masses and the pattern of supersymmetry breaking.

\vspace{0.4cm}

We would like to conclude with some outlook on supersymmetry breaking
via gaugino condensation in the type of models considered in this paper.
Here, we have only analyzed the simplest case of gaugino condensation
in the $E_8$ sectors. As we have seen, one expects additional hidden
sectors in the effective theory that originate from the degrees of freedom on
the five--brane worldvolumes. In particular, under certain
circumstances~\cite{np}, these hidden sectors have non--Abelian gauge
groups. Hence, gaugino condensation might also occur in these sectors.
Due to the origin of these new gauge groups, their gauge kinetic
functions might be quite different from the ``conventional'' ones in
the $E_8$ sectors. This might shed new light on the generic problems
of gaugino condensation~\cite{kl1}, such as the stabilization of the
dilaton at large field values. We will address these issues in a
future publication~\cite{inpr}.

\vspace{0.4cm}

{\bf Acknowledgments} 
A.~L.~would like to thank Graham Ross and Andrea Romanino for discussions.
A.~L.~is supported by the European Community under contract No. 
FMRXCT 960090. B.~A.~O.~is supported in part by 
DOE under contract No. DE-AC02-76-ER-03071. D.~W.~is supported in part by
DOE under contract No. DE-FG02-91ER40671. 
 


\end{document}